\documentclass[%
 reprint,
 amsmath,amssymb,
 aps,
]{revtex4-2}

\usepackage{graphicx}
\usepackage{dcolumn}
\usepackage{bm}
\usepackage{orcidlink}

\usepackage{xcolor}

\def\lsim{\;\raise0.3ex\hbox{$<$\kern-0.75em\raise-1.1ex\hbox{$\sim$}}\;}
\def\gsim{\;\raise0.3ex\hbox{$>$\kern-0.75em\raise-1.1ex\hbox{$\sim$}}\;}
\def\lsim{\;\raise0.3ex\hbox{$<$\kern-0.75em\raise-1.1ex\hbox{$\sim$}}\;}
\def\gsim{\;\raise0.3ex\hbox{$>$\kern-0.75em\raise-1.1ex\hbox{$\sim$}}\;}
\def\Msun{M_\odot}

\def\cmc{\rm ~cm^{-3}}

\def\cmc{\rm ~cm^{-3}}
\def\diff{\rm ~cm^2~s^{-1}}

\def\ergs{\rm ~erg~s^{-1}}

\def\enf{\rm ~erg~cm^{-2}~s^{-1}}

\def\ecsb{erg cm$^{-2}$ s$^{-1}$ arcsec$^{-2}$ }
\def\ecsb2{erg cm$^{-2}$ s$^{-1}$ arcsec$^{-2}$}

\usepackage{url}
\usepackage{xcolor}
\usepackage{ulem}
\definecolor{newcolor}{rgb}{.8,.349,.1}

\begin{document}
\preprint{APS/123-QED}

\title{ PeV particle acceleration and non-thermal emission in the `minimalist' model of the extended jets in W50/SS433}

\author{Andrei.M. Bykov\,\orcidlink{0000-0003-0037-2288}}
\email{byk@astro.ioffe.ru}
\affiliation{Ioffe Institute, 194021, Saint-Petersburg, Russia}
\author{Sergei.M. Osipov\,\orcidlink{0000-0001-8806-0259}}
\email{osm.astro@mail.ioffe.ru}
\affiliation{Ioffe Institute, 194021, Saint-Petersburg, Russia}
\author{Vadim.I. Romansky\,\orcidlink{0000-0003-1863-2957}}
\email{romanskyvadim.astro@mail.ioffe.ru}
\affiliation{Ioffe Institute, 194021, Saint-Petersburg, Russia}
\author{Yury.A. Uvarov \,\orcidlink{0000-0002-4962-5437}}
\email{uv@astro.ioffe.ru}
\affiliation{Ioffe Institute, 194021, Saint-Petersburg, Russia}
\author{Eugene Churazov\,\orcidlink{0000-0002-0322-884X}}
\email{churazov@mpa-garching.mpg.de}
\affiliation{Max Planck Institute for Astrophysics, Karl-Schwarzschild-Str. 1, D-85741 Garching, Germany}\
\affiliation{Space Research Institute (IKI), Profsoyuznaya 84/32, Moscow 117997, Russia}
\author{Ildar Khabibullin\,\orcidlink{0000-0003-3701-5882}}
\email{ildar@mpa-garching.mpg.de}
\affiliation{Universitäts-Sternwarte, Fakultät für Physik, Ludwig-Maximilians-Universität München, Scheinerstr.1, 81679 München, Germany}
\affiliation{Max Planck Institute for Astrophysics, Karl-Schwarzschild-Str. 1, D-85741 Garching, Germany}\
\affiliation{Space Research Institute (IKI), Profsoyuznaya 84/32, Moscow 117997, Russia}
\date{\today}

\begin{abstract}
The W50 nebula around microquasar SS~433, powered by supercritical accretion, features two `extended jets' (tens of pc long and a few pc wide) from which polarized X-ray and very high energy radiation above 100 TeV is detected. Here we present a model of very high energy particle acceleration in these extended jets. In the `minimalist' model (discussed in Churazov, Khabibullin, and Bykov, 2024), a collimated outflow aligned with the rotation axis is propagating through a more isotropic wind produced by the accretion disk. The observed extended X-ray jets with bright knots in this model are associated with the formation of strong recollimation MHD shocks after the collision of the collimated outflow with the isotropic wind termination surface. 
The spectra of electrons and protons up to PeV energies are simulated with a nonlinear Monte Carlo model of diffusive shock acceleration with turbulent magnetic field amplification. The overall efficiency of the jets power transfer to accelerated protons in this model is above 10\%  and about 0.5\% for electrons above 50 TeV. The magnetic field amplification by Bell's instability due to the electric current of cosmic rays escaping the accelerator produces highly anisotropic magnetic turbulence in the shock downstream. This results in the polarized synchrotron X-ray emission with the photon electric vector predominantly transverse to the jet direction and the degree of polarization above 20\%. The model is able to reproduce the observed spectra and intensity profiles of non-thermal X-ray and gamma-ray emission, which are both dominated by the leptonic radiation.     
\end{abstract}

\maketitle


\section{Introduction}

W50 is an extended radio-bright nebula that is powered by a compact source SS~433, which is a
binary system with an accreting black hole (or neutron star) \citep[see, e.g.][for a review]{1984ARA&A..22..507M,2004ASPRv..12....1F,2025arXiv250601106C}. This system operates in a hyper-accretion regime, which is much less studied
than sub-Eddington sources \citep[e.g.][]{2015NatPh..11..551F}. The accretion in this system can, for a long time, be sustained at the highly super-Eddington rate  \citep{2017MNRAS.471.4256V,2023NewA..10302060C}.
The total energy that is injected by such a source into the surrounding medium is comparable to the energy output of a supernova explosion. So SS~433 can strongly affect the Interstellar Medium (ISM) in its vicinity, producing and powering an extended ($\gtrsim 100$ pc) nebula  \citep[e.g.][]{1980ApJ...238..722B}.

During hyper-accretion, SS~433 produces trans-relativistic narrow baryonic jets with bulk velocity $\sim 0.26\,c$ and an opening angle of less than 2 degrees \citep[e.g. ][]{1980ApJ...238..722B,1981Ap&SS..79..387C}. This is a
prominent feature of SS~433.
These narrow jets are seen in X-ray and optical spectra as pairs of blue- and red-shifted lines, which for X-ray and optical lines come, respectively, from the regions located within $\lesssim 10^{12}$ cm and $\lesssim 10^{15}$ cm from the compact source \citep[][]{2002ApJ...564..941M,2004ASPRv..12....1F,2016MNRAS.455.1414K,2018ApJ...867L..25B,2019AstL...45..299M,2019A&A...624A.127W}. 
However, the baryonic jets remain invisible beyond the distance of $\sim0.1$ pc from the central source \citep[e.g. ][]{1981ApJ...246L.141H,2004ApJ...616L.159B,2005MNRAS.358..860M,2008ApJ...682.1141M}. 
Since at large scales  (40 to 110 pc)  the surrounding SS~433  radio- and ${\rm H}\alpha$-bright nebula W50 has an elongation aspect ratio 3:1 along the jets precession axis, it has been considered that the kinetic energy of the jets powers the W50 nebula \citep[][]{1980ApJ...238..722B,1980MNRAS.192..731Z,1983ApJ...272...48E,1993ApJ...417..170P,2000A&A...362..780V,2008MNRAS.387..839Z,2011MNRAS.414.2838G,2014ApJ...789...79A,2015A&A...574A.143M,2017A&A...599A..77P,2021ApJ...910..149O}. 

There is no direct evidence of the termination of the narrow jets in the radio, optical and X-ray maps of W50 \citep[][]{1998AJ....116.1842D,2011MNRAS.414.2838G,2017MNRAS.467.4777F,2018MNRAS.475.5360B}, while a
broader and more extended X-ray structures are seen on both sides of SS~433  aligned with the narrow optical jet precession axis \cite[e.g.,][]{1983ApJ...273..688W,1994PASJ...46L.109Y,Safi-Harb1997,1999ApJ...512..784S,Brinkmann2007,Safi-Harb2022}. These extended X-ray jets appear at a distance of $\sim 20\,$pc from the compact source and have an opening angle of $\sim 20^\circ$ which is twice smaller than the precession amplitude of the narrow jets.
The X-ray structures are probably of synchrotron origin \citep[e.g.,][]{Brinkmann2007}, which is confirmed by the detection of X-ray polarization by the IXPE telescope \citep{IXPE2024SS433}. The very high energy (VHE) gamma-ray emission from the source was predicted in a number of models  \citep[see, e.g.,][]{Safi-Harb1997, Reynoso2008, 1998NewAR..42..579A, 
2020ApJ...904..188K, 2020ApJ...889..146S}. Recently, the extended very high energy gamma-ray emission was detected from W50 \citep[][]{2018Natur.562...82A,HESS2024SS433,LHAASO2024SS433}, indicating the acceleration of PeV-regime particles in the source. The physical origin of these extended structures and their connection to the narrow transrelativistic jets currently launched by the system remains unclear \citep[e.g. ][ and references therein]{2020MNRAS.495L..51C}

 The `minimalist' model \citep{Churazov} explains the extended jet as a result of the powerful two-component anisotropic outflow produced by a supercritical accretion disk. In this model, a more collimated outflow (polar wind), aligned with the rotation axis of the binary, coexists with a more isotropic wind. The extended X-ray jets observed are associated with the recollimation shocks, which appear right after the collision of the collimated outflow with the isotropic wind termination surface.  The structure of  MHD outflow in the interaction region of the collimated wind with the termination surface of the isotropic wind was simulated with the PLUTO MHD code \cite{Mignone2007}. We present a model of very high-energy particle acceleration in the extended jets of the SS~433/W50 system following the two-component outflow scenario.

The results of MHD modeling are described in Section \ref{sec:morphological}. Using the spatial distributions of the accelerated particles and magnetic field spectra obtained in the model, we construct profiles and spectra of non-thermal X-ray and gamma radiation of this object.
The modeling includes the following steps, split across several sections.

Section \ref{sec:MC}. Based on the plane-parallel Monte Carlo model of particle acceleration and magnification of magnetic fields near the shock fronts, the spectra of accelerated particles and magnetic fields near the shock waves obtained in the MHD model are found.

Section \ref{sec:CorSh}. Based on MHD modeling of the passage of turbulence with significant density fluctuations through the shock front, profiles of the longitudinal $B_{\parallel}$  and transverse $B_{\perp}$ components of the magnetic field behind the shock front are obtained using the PLUTO code.

Section \ref{sec:Spectrum}. Next, the spectra and profiles of the X-ray synchrotron emission are calculated based on the spectra of accelerated electrons and magnetic field profiles. Gamma Rays in the TeV range are also calculated. The simulation parameters are selected to match the model with the observational data \citep[][]{Brinkmann2007, Safi-Harb2022, HESS2024SS433, LHAASO2024SS433}.
The anisotropy of the magnetic field in the shock downstream leads to the observed polarization of the X-ray radiation. The X-ray polarization obtained in the model near the shock wave is compared with the observational data from the IXPE observatory \citep{IXPE2024SS433}.

High energy particle acceleration in microquasars as a potential contributor to the galactic cosmic ray pool was widely discussed (see e.g. \cite{2002A&A...390..751H, 2005A&A...432..609B,2017SSRv..207....5R}), while recently the interest to SS 433 as a pevatron was raised in \cite{Churazov,2024arXiv241108762P,2025arXiv250620193Z,2025arXiv250622550C,2025arXiv250721048W} by VHE gamma-ray observations reported in \cite{HESS2024SS433,LHAASO2024SS433}.

\section{MHD modeling of the morphological structure of  SS433/W50}
\label{sec:morphological}

\begin{figure*}
\includegraphics[scale=0.6]{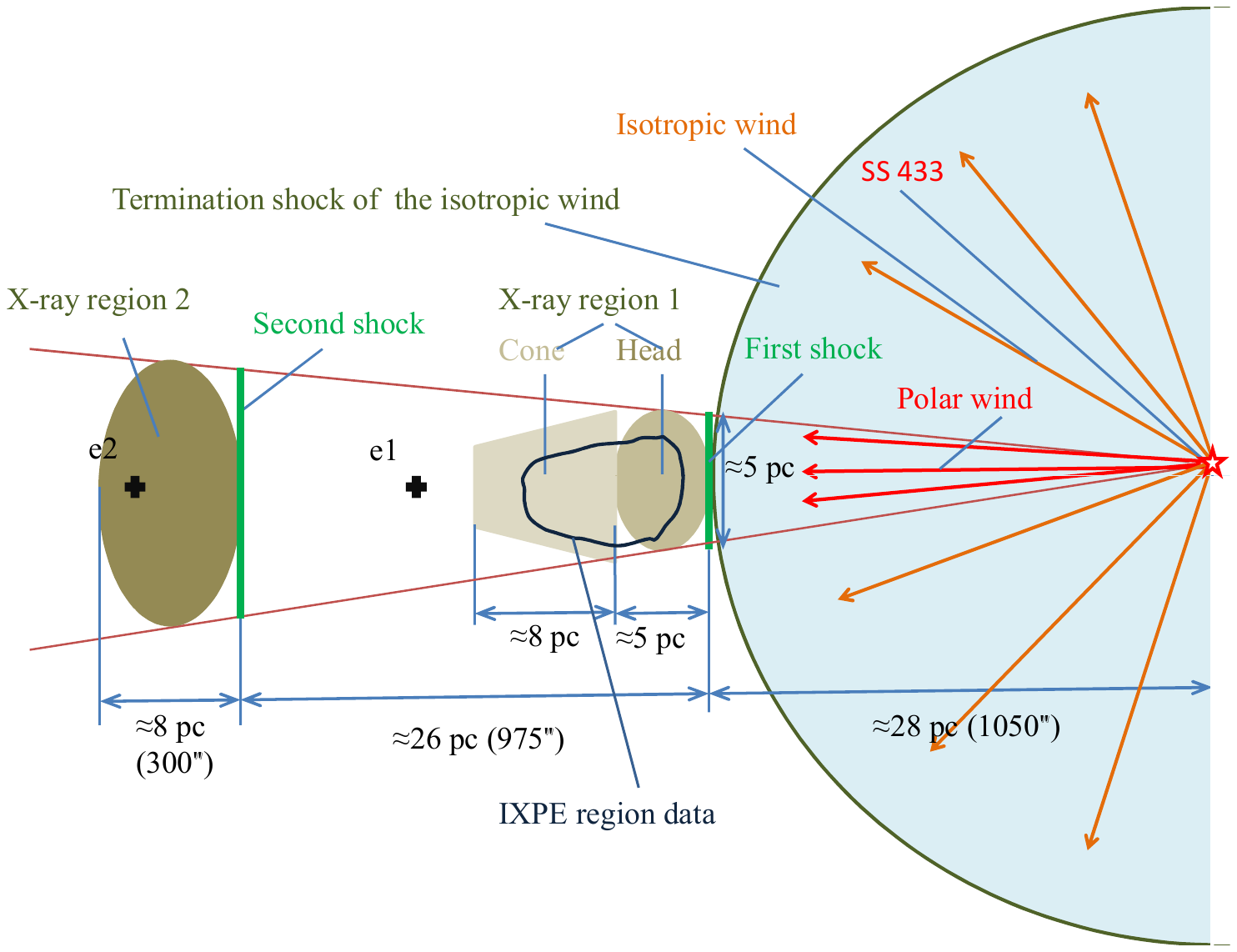}
\caption{Schematic representation of the SS433+W50 system adopted here. The structures with bright non-thermal X-ray emission in the extended jet of SS 433 which we model here are marked as  `Cone", `Head", `e1' and `e2". They were detected and discussed in the papers \cite{Safi-Harb1997,Brinkmann2007,Safi-Harb2022}.}
\label{Scheme}
\end{figure*}

In this section, we perform axisymmetric MHD modeling of the W50 nebula around the microquasar SS 433. It has a characteristic shape: a central spherical part and two opposite lobes to the east and west of the central source. The X-ray emission of the nebula is very inhomogeneous, with several bright X-ray regions - `e1", `e2", and `e3' to the east and `w1' and `w2' to the west \cite{Safi-Harb1997}. Following \cite{Churazov}, we model a wind from the central microquasar with two components - quasi-isotropic slow wind and a more collimated outflow, the polar wind. The kinetic power of each component is about $10^{39}$ erg/s, and the velocities are $\approx3000$ km/s for the isotropic wind and $\approx0.2~c$ for the jet. Interactions of the collimated outflow, isotropic wind, and ambient medium cause several recollimation shocks in the jet, which explains the existence of different X-ray emitting regions. The sketch of the hydrodynamical model with the emission regions shown from \cite{Safi-Harb2022, IXPE2024SS433} is presented in Figure \ref{Scheme}.

We perform numerical simulations of the W50 nebula using the MHD code PLUTO \cite{Derouillat}. We used a 2D setup in cylindrical coordinates. In the central region, at the sphere with radius of 5 pc, we set internal boundary conditions with fixed density, pressure, and radial velocity, depending on the polar angle $\theta$ of the current grid point. For small angles, the outflow parameters correspond to the trans-relativistic collimated outflow, and for the larger angles to the isotropic wind. The ambient medium has constant density and pressure. 

\begin{figure}
\includegraphics[scale=0.36]{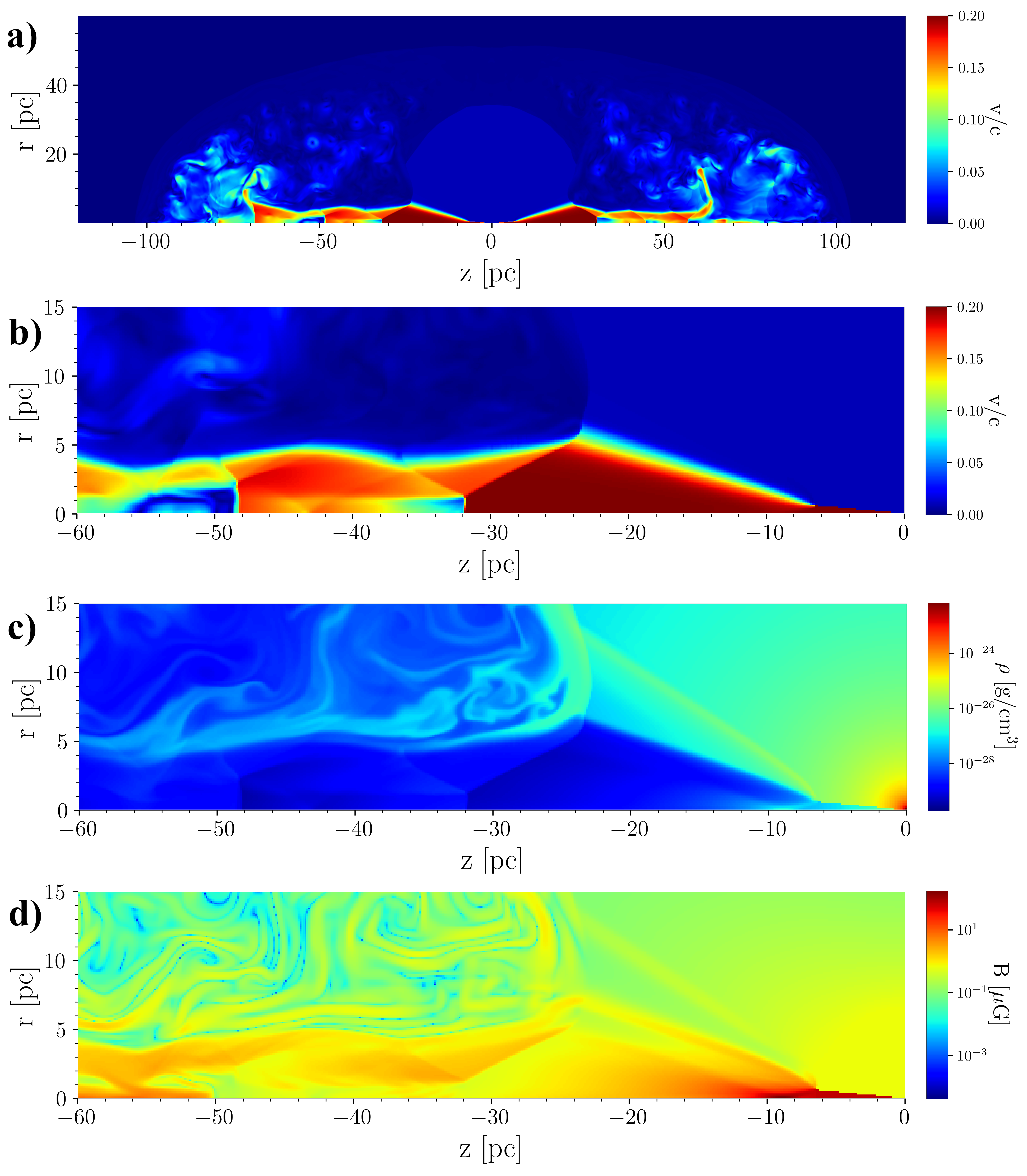}
\caption{Morphology of the W50 nebula in simulations: a) large-scale velocity profile in the entire computational domain, b) velocity near e1 and e2 regions, c) density near e1 and e2 regions, d) magnetic field near e1 and e2 regions.} 
\label{velocity}
\end{figure}

The magnetic field was set to 100 $\mu$G in the jet base and 1 $\mu$G in the ambient medium and isotropic wind, and directed along the z-axis. We managed to obtain an elongated structure with two strong recollimation shocks with noticeable Mach diamonds, located approximately at distances 32 and 48 pc from the central source, close to observed bright X-ray emitting regions, named `e1' and `e2' in the eastern jet \cite{Safi-Harb1997}. The structure of hydrodynamical flow is shown in Fig.~\ref{velocity} - large-scale velocity map in the top panel, and zoomed maps of velocity, density, and magnetic field in the region between the central source and `e2' in the lower panels. The parameters for this setup are: kinetic power of the isotropic wind $P_i$ is equal to kinetic power of the collimated outflow $P_j = 2.1\times10^{39}$ erg/s, the velocity of isotropic wind is $v_i=3000$ km/s, the velocity of collimated outflow is $v_j=0.2~c$, its half-width angle is $5^{\circ}$, the ambient number density is $\rho_{amb} = 0.005~\rm{cm^{-3}}$ and the temperature $T_{amb} = 8\times10^4~\rm{K}$. We assumed that the accretion disk outflows which produced the extended X-ray jet are propagating through a nebula of low plasma density created previously  by the supernova event and modified by the winds from the binary (see e.g. a model of the nebula in \citep{2021ApJ...910..149O}). 

\begin{figure}
\includegraphics[scale=0.36]{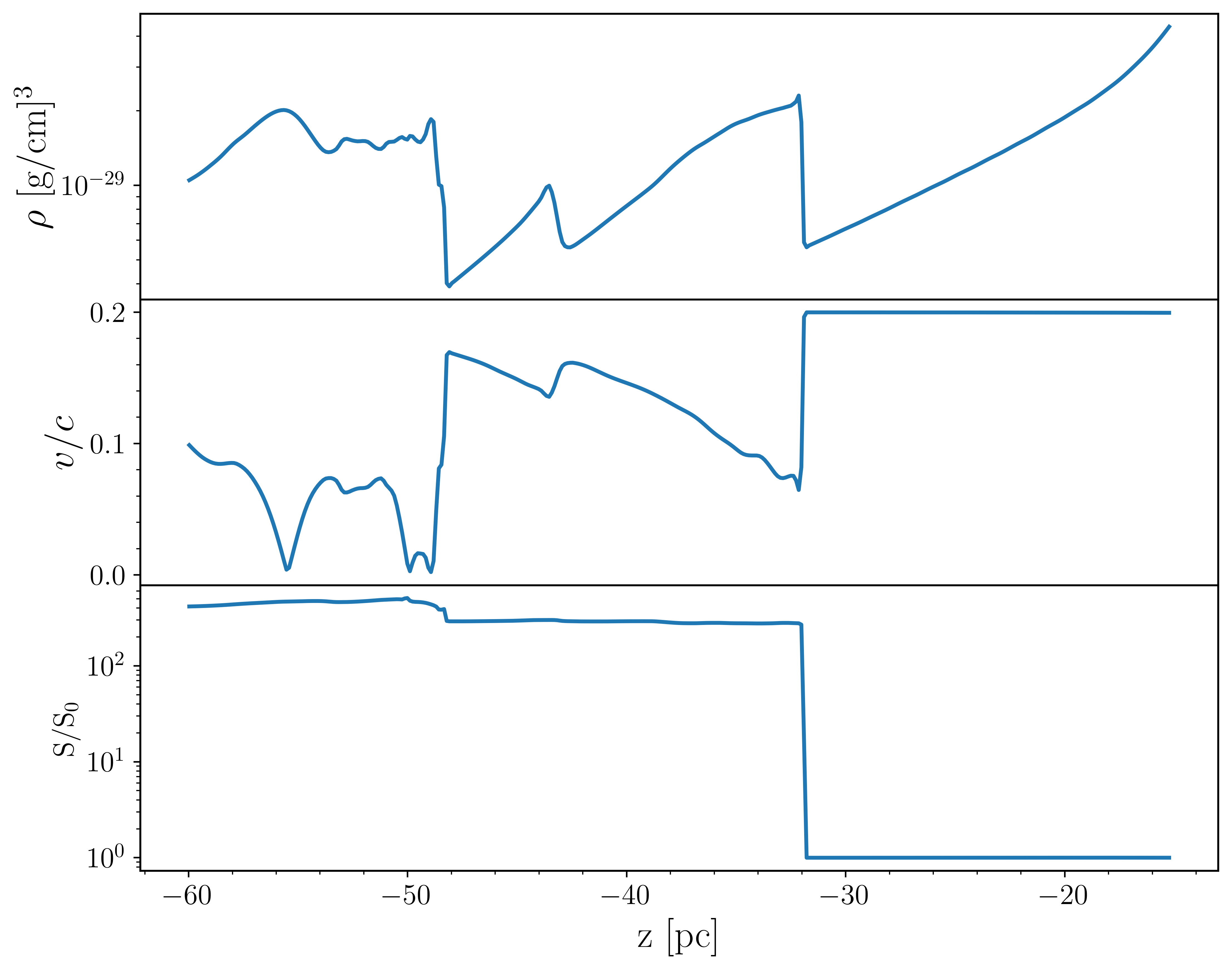}
\caption{Density, velocity, and entropy profiles in the MHD run shown in Fig.~\ref{velocity}.} 
\label{MHD_profile}
\end{figure}

 The ambient matter parameters used above are not unique. It is possible to obtain a similar structure of the extended X-ray jets and nebulae with higher ambient density and lower temperature e.g.  ($\rho_{amb}=0.05~\rm{cm^{-3}}$, $T_{amb} = 8.6\times10^3~\rm{K}$). In this case, the structure  with strong internal shocks,  similar to that shown in Fig. \ref{MHD_profile} appears at later times of about 100, 000 years. We left the detailed study of the parameter space of the ambient matter allowed by multiwavelength observations of W50 for future study, and here we model the VHE particle acceleration and nonthermal emission of the extended X-ray jets using the main setup described above.

The first recollimation shock occurs close to the termination shock of the isotropic wind. The later behavior of the outflow - will it propagate further with a series of recollimation shocks, or stop after the first one, is determined by the pressure of the ambient medium. It should be low enough for the collimated outflow to propagate a large distance. In addition, relatively low ambient pressure is necessary to form Mach shocks with plasma flow normal to the shock front, instead of a system of crossing oblique shocks, as shown in \cite{Marti2018}. In numerical simulations, spatial resolution should be high enough to resolve Mach diamonds.

The one-dimensional  density, velocity, and entropy profiles of the hydrodynamic flow near the regions `e1' and `e2' are shown in Figure~\ref{MHD_profile}. The density jumps at the two shocks are close to 4, which is a sign of the strong hydrodynamic shocks. Entropy jumps are relatively large as well. Also, there is a weak shock with smaller jumps between two strong shocks. Kinetic Monte Carlo collisionless shock models, which take into account the energy fluxes of accelerated particles, showed that the compression at the shock front can significantly exceed the adiabatic value 4 (see e.g. simulation in \cite{Bykov3inst2014}).

In the minimalists' scenario, the main goal was reproducing the overall morphology of the W50 radio emission and the emergence of the `extended X-ray jets' as a result of recollimation shocks in the anisotropic wind scenario. This model has several free parameters, including overall energetics, age of the system, density of the ambient medium, velocities and densities of the isotropic and polar winds, and the opening angle of the polar component. Various combinations of these parameters can lead to a qualitatively similar morphology of the nebula (see Fig.A1 in \cite{Churazov}). Here, our primary goal is to find parameters that lead to the efficient acceleration of the particles that give rise to the observed emission of EXJs. Knowing that accelerating electrons to $\gsim$ 100 TeV energies is especially efficient for strong shocks with velocities $\gtrsim$ 0.2c \cite{Churazov}, we ran the MHD version of a set-up with this value of the polar component velocity, but did not require the morphology of the simulated nebula boundary to match the observed one perfectly. We leave the task of finding a parameter combination that satisfies all observational requirements best for a future study.

\section{Monte Carlo model of particle acceleration and amplification of magnetic fields}
\label{sec:MC}

In this section, we simulate particle acceleration and magnetic amplification using the stationary plane-parallel  Monte Carlo code for diffusive shock acceleration \citep[][]{Bykov3inst2014,Bykov2022Univ}. The parameters of the background plasma in the shock upstream are taken from the MHD   model described in  Section \ref{sec:morphological}.  In the Monte Carlo code, the conservation laws of energy and momentum at the shock front are achieved in an iterative process.  In this study, the Monte Carlo model takes into account the magnetic field amplification by Bell's instability in the upstream caused by the accelerated particle current relative to the background plasma. In the Monte Carlo model, the free escape boundary (FEB) of the particles is located upstream ($L_{FEB}$ is the distance from the shock front to the FEB). We assume that the FEB should be smaller than the transverse size of the observed area. Thus, we take $L_{FEB}=2$ pc and $L_{FEB}=4$ pc, respectively, for the first and second shocks (see Figure \ref{Scheme}). The following parameters are set in the Monte Carlo simulation: $u_{sh}=0.2c$, $n_{0}=5\cdot 10^{-6}$~cm$^{-3}$, $T_{0}=5\cdot 10^{5}$~K  for the first shock and  $u_{sh}=0.18c$, $n_{0}=3\cdot 10^{-6}$~cm$^{-3}$, $T_{0}=10^{8}$~K for second shock, where $u_{sh}$ is the velocity of the background plasma in the far unperturbed upstream, $n_{0}$ the number density of protons in the far upstream, $T_{0}$ is the background plasma temperature at the FEB (see Section \ref{sec:morphological}).  For both shocks, $B_{0}=0.2$~$\mu$G, $B_{turb0}=1.0$~$\mu$G, where $B_{0}$ is the uniform magnetic field at the FEB and $B_{turb0}$ is the rms turbulent magnetic field at the FEB, respectively. The turbulent magnetic field at the FEB has the Kolmogorov spectrum with an energy-carrying scale of $L_{ls}\approx 3$ pc.

\begin{figure}
\includegraphics[scale=0.36]{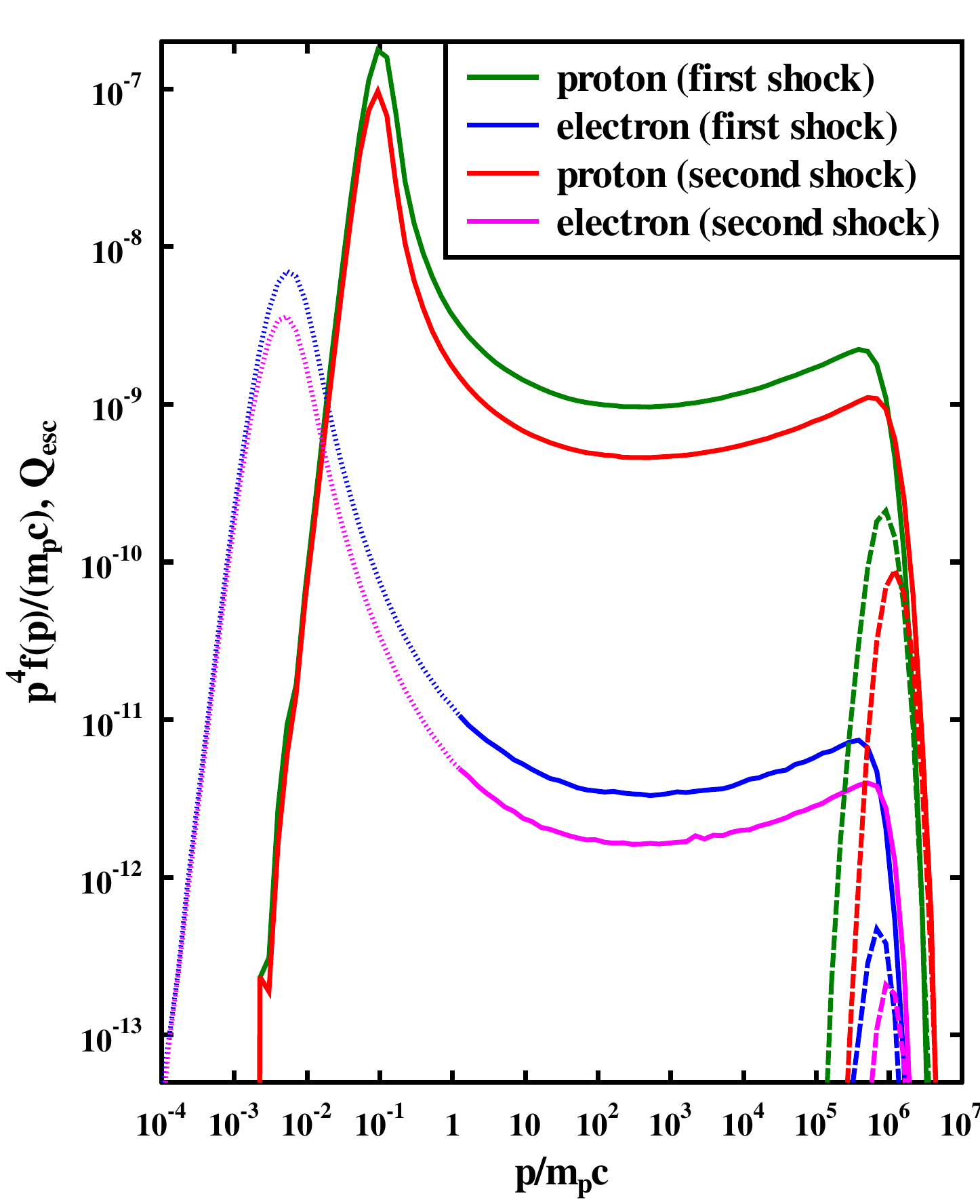}
\caption{Simulated Monte Carlo spectra of accelerated particles at the shock front position (derived in the shock rest frame) are shown in solid curves. Dashed curves show the flux of accelerated particles that escape through FEB into the far upstream. Green and red curves are the spectra of accelerated protons at the first and the second shock, respectively. Blue and magenta curves show the spectra of accelerated electrons at the first and the second shock, respectively. Dotted curves show the assumed low-energy extrapolation of the electron spectra down to the thermal pool. These curves are shown only for illustration and are not used in simulations of non-thermal high energy radiation.} 
\label{pdf_shock_Q_esc}
\end{figure}

The simulations provided the rms magnetic field is $B_{rms}\approx15\ \mu G$ and $B_{rms}\approx9\ \mu G$, and the flow velocities $u_{2}=10.3\cdot 10^{8}$ cm/s and $u_{2}=9.3\cdot 10^{8}$ cm/s directly behind the front of the first and second shocks, respectively. The spectra of the accelerated particles at the shock front for both shocks are shown in Fig.~\ref{pdf_shock_Q_esc}. The flux of accelerated particles escaping through FEB in the far upstream is
\begin{equation}
\label{Qesc}
    Q_{esc}=\frac{j_{p}^{cr}(L_{FEB},p)p^{4}}{4\pi m_{p}c^{2}},
\end{equation}
\begin{equation}
\label{jcr}
   j^{cr}(z)=\int_{0}^{\infty} j_{p}^{cr}(z,p)p^{2}dp,
\end{equation}
where $p$ is the particle momentum, $m_{p}$ is the mass proton, $j^{cr}(z)$ is the current of accelerated particles at a given point. The particle distribution functions are normalized as follows
\begin{equation}
\label{fcrnorm}
   n(z)=4\pi\int_{0}^{\infty} f(z,p)p^{2}dp,
\end{equation}
where $n(z)$ is the accelerated particle number density at the point z.

The nonlinear Monte Carlo model used here obeys the momentum and energy conservation laws and simulates the proton spectra within the basic assumptions of the proton scattering prescription for all of the protons without an injection mechanism. The electrons are the minor component in this case and one can expect that their spectral shape will reproduce that of the accelerated protons starting from some energy well above $T_p$, actually $\gsim m_p c^2$, where both electrons and protons are relativistic and are scattered by the magnetic turbulence depending on their rigidities, i.e. are in DSA regime. At low energies, the process of electron energization is different from that of protons (see e.g. \cite{2015PhRvL.114h5003P}). To be injected in DSA Fermi acceleration, electrons are subject to the shock drift acceleration, and it takes many cyclotron cycles for them to enter the DSA regime \cite{MarcowithEtal2016}. The available particle-in-cell simulations allowed to estimate the electron-proton number ratio measured at a given energy in accelerated particles for different shock velocities, which was found typically to be below 10$^{-2}$ (see e.g. \cite{2015PhRvL.114h5003P,Bykov2022Univ,2024ApJ...968...17G}). In our case of sub-relativistic shock in the outflow with weak magnetic field, we used the ratio 3.3$ \times 10^{-3}$ to fit the observed level of synchrotron radiation, as it is shown in Section \ref{sec:Spectrum}. The Monte Carlo simulation of electron acceleration took into account the synchrotron-Compton energy losses of relativistic electrons, which limit the maximal energy of accelerated electrons.

\section{MHD simulations of the passage of turbulence through a shock front}
\label{sec:CorSh}

When MHD turbulence containing strong density fluctuations passes through the shock,  strong oscillations of the shock front occur. These oscillations lead to significant turbulent fluctuations in the downstream flow velocity. Moreover, turbulent velocity fluctuations amplify the magnetic field downstream by the mechanism of a small-scale dynamo. It appears that in such a system, the longitudinal component of the magnetic field relative to the shock normal is amplified more strongly than the transverse components \citep{inoue13,Zirakashvili_2008pol,lazarian_MFA22,federath23,2024PhRvD.110b3041B}. Thus, there is a region downstream where the longitudinal component of the turbulent magnetic field is greater than the transverse one. Synchrotron radiation of electrons from this region should be highly polarized with polarization direction transverse to the shock normal.

We simulate the passage of MHD turbulence amplified by the Bell instability through the shock to determine the anisotropy of the magnetic field downstream in a way similar to that in the paper \citep{2024PhRvD.110b3041B}. The 3D calculations are performed on the basis of the MHD module of the open-source code PLUTO. The calculation is performed in two stages: in the first stage, the amplification of MHD turbulence by the Bell instability is simulated in a cubic box (512x512x512 cells), and in the second stage, the passage of turbulence through the shock front is simulated (see \citep{2024PhRvD.110b3041B}). The sizes of the simulation box at the first stage in all directions vary from $-l_{\ast}$ to $l_{\ast}$, where $l_{\ast}$ is the normalization length of the PLUTO code. The size of the simulation box at the second stage along the z-axis is from 0 to $10l_{\ast}$, and along the x and y axes is from $-l_{\ast}$ to $l_{\ast}$ (512x512x2560 cells). The box of the first stage is used at the second stage as a boundary condition on the right boundary along the z-axis. The boundary condition on the left boundary of the z-axis is an outflow. The boundary conditions along the x and y axes are periodic. We vary the length of $l_{\ast}$ to achieve agreement with the observational data. The energy-carrying scale of turbulence $l_{0}$ (calculated at the first stage) passing the shock is approximately equal to $l_{\ast}/3$. At the initial moment, the shock is set perpendicular to the z-axis at the point $z=2 l_{\ast}$ and remains at rest in the absence of perturbations.

The simulation parameters in the first stage of the calculations are taken in accordance with the conditions upstream of the first shock described in Section \ref{sec:MC}. The ratio of the rms deviation of the density to the average density is $\delta \rho/\rho\approx 0.6$ at the end of the first-stage simulation.

\begin{figure}[!th]
\includegraphics[bb= 30 10 570 570,clip, scale=0.44]{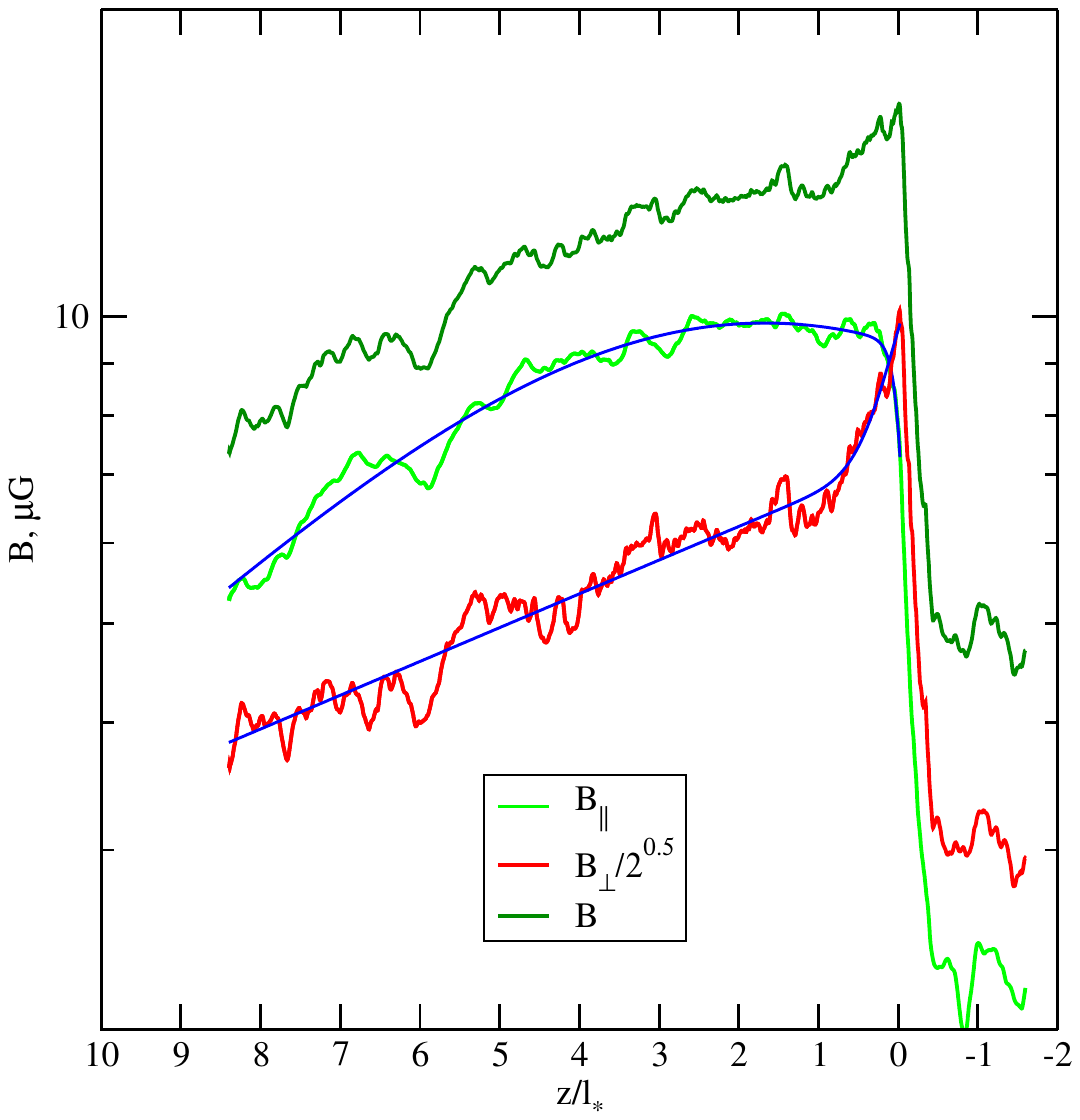}
\caption{Profiles of RMS components of the magnetic field in the shock downstream. The turbulence is produced by the passage of fluctuations amplified in the upstream by Bell's instability of accelerated cosmic rays. The green curve is the parallel magnetic field component. The red curve is the perpendicular magnetic field component. The dark green curve is the total magnetic field. Blue curves are fits to the data.} 
\label{B2_Bell_art}
\end{figure}

\begin{figure}[!th]
\includegraphics[bb= 30 10 570 570,clip, scale=0.44]{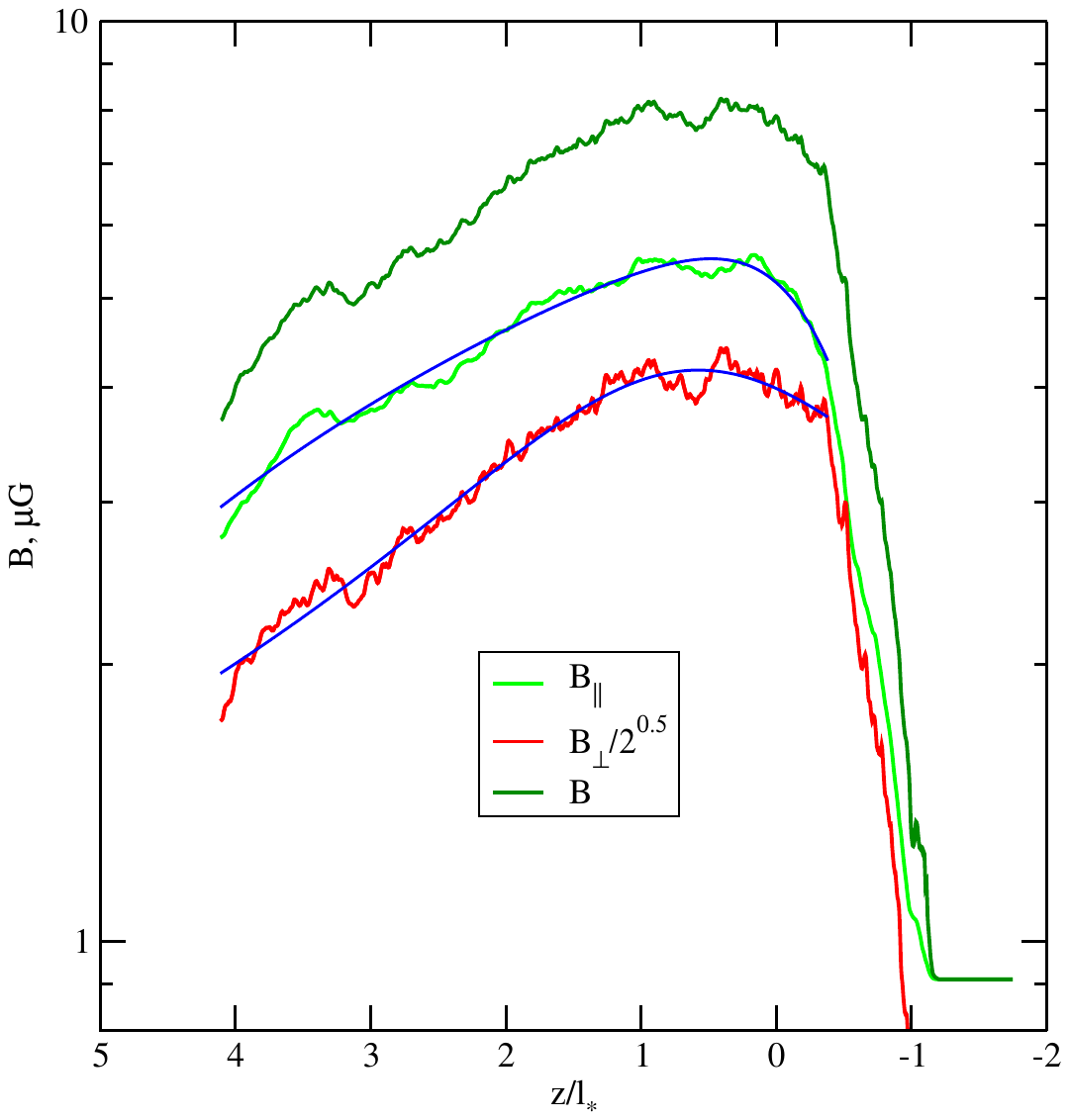}
\caption{Profiles of RMS components of the magnetic field in the shock downstream produced by the shock amplification of the preexisting long-wavelength fluctuations in the jet. The green curve is the parallel magnetic field component. The red curve is the perpendicular magnetic field component. The dark green curve is the total magnetic field. Blue curves are fits to the data.} 
\label{B2_all_longscale_art}
\end{figure}

The results of the simulations for turbulence with strong density fluctuations and magnetic fields amplified by the Bell instability are shown in  Fig.~\ref{B2_Bell_art}. The origin of the z coordinate in Fig.~\ref{B2_Bell_art} and Fig.~\ref{B2_all_longscale_art} is shifted to the location of the shock front. These figures show the profiles of the magnetic field components averaged over slices with fixed coordinate z. In the figures, $B_{\parallel}$ is the RMS of the magnetic field component 
parallel to the shock normal of the undisturbed shock. $B_{\perp}=\sqrt{B_{x}^{2}+B_{y}^{2}}$ is the RMS perpendicular component of the magnetic field, where $B_{x}$ is the RMS value of the x projection of the magnetic field, $B_{y}$ is the RMS value of the y projection of the magnetic field, and $B=\sqrt{B_{\parallel}^{2}+B_{\perp}^{2}}$. The blue curves in the figures show the same data after filtering out small-scale fluctuations. They are used in subsequent calculations.

The nonlinear DSA Monte Carlo model derives the magnetic field fluctuation spectra but does not provide the polarization of the magnetic fluctuations needed to simulate a polarization of X-ray synchrotron radiation. To get the turbulent magnetic field polarization, we used a combined model, where the Monte Carlo simulations provided the electric current of CRs that were accelerated at the shock and left the accelerator. The turbulence produced by Bell's instability in the shock vicinity  simulated with the MHD code PLUTO once the CR current in the shock upstream (taken from the Monte Carlo model) is fixed. This allows us then to model the polarization of the synchrotron X-ray emission (see for details \citep{2024PhRvD.110b3041B}).

The system has a broad range of scales. We assumed here that the plasma flowing into the shock from the far upstream has the energy-carrying scale of turbulent fluctuations of the order of the collimated flow width. Fluctuations of longer wavelength just result in a displacement of the shock position. The characteristic scale of the fluctuations amplified by Bell's instability, which is driven by the electric current of accelerated particles leaving the accelerator, is substantially smaller than that in the pre-existing jet turbulence. The passage of turbulence with the two very different energy-carrying scales through the shock front is simulated in two steps. First, with the PLUTO code, we model the MHD evolution of the cosmic-ray driven turbulence, which is accompanied by diffusive shock acceleration. High energy protons accelerated by strong shocks penetrate upstream, and their current initiates Bell's turbulence there. The Cosmic Ray current in the upstream region, which is a fixed value used in the PLUTO MHD simulation, is obtained with kinetic Monte Carlo modeling of the diffusive shock acceleration. Magnetic field, velocity, and density fluctuations, produced due to the non-linear evolution of Bell's turbulence passed through the MHD shock, form the turbulence in the shock downstream. This setup is similar to the one described in \cite{2024PhRvD.110b3041B}. The energy-containing scale of Bell's turbulence is of the order of the gyro-radius of the highest-energy accelerated particles. We assume that the scale separation between the cosmic-ray-driven and the pre-existing turbulence in the jet is large. Therefore, in Bell's turbulence simulation, the jet flow is homogeneous far upstream, and its magnetic field directed along the jet has a magnitude of 1 $\mu$G. Then, we simulated the passage of the preexisting jet plasma turbulence with a much larger scale than that of the cosmic-ray driven turbulence, for which the energy-carrying scale is of the order of the jet width. This is needed to estimate the magnetic field magnitude at the distances from the shock which are larger than the dissipation length of Bell's turbulence (see Fig. \ref{B2_Bell_art}). The shock-amplified preexisting turbulence dominates the magnetic field in the downstream after the cosmic ray turbulence dissipation.

Further, we simulated the synchrotron radiation of shock-accelerated electrons in the turbulent magnetic fields downstream to model the spectra, intensity profiles, and polarization of X-ray emission.

\section{X-Ray emission simulation}
\label{sec:Spectrum}

X-ray emission from SS433/W50 nebula has several specific features - inhomogeneous spatial profile with bright regions, energy spectra, and polarization. In this section, we present a model describing observed X-ray emission from the eastern lobe of W50 \citep{Brinkmann2007,Safi-Harb2022}, via synchrotron radiation of electrons, accelerated on two trans-relativistic shocks, in the inhomogeneous anisotropic turbulent magnetic field. Assumption of inhomogeneous profile of turbulent magnetic field is supported by the fact that narrow peaks in observed profiles in the `Head' region in energy ranges 0.3-10 keV and 3-30 keV \cite{Safi-Harb2022} have similar widths, which can be explained by the structure of the magnetic field. On the contrary, under the assumption that the peak width is a result of synchrotron losses,  peaks should have different widths.

\begin{figure}[th!]
\includegraphics[scale=0.65]{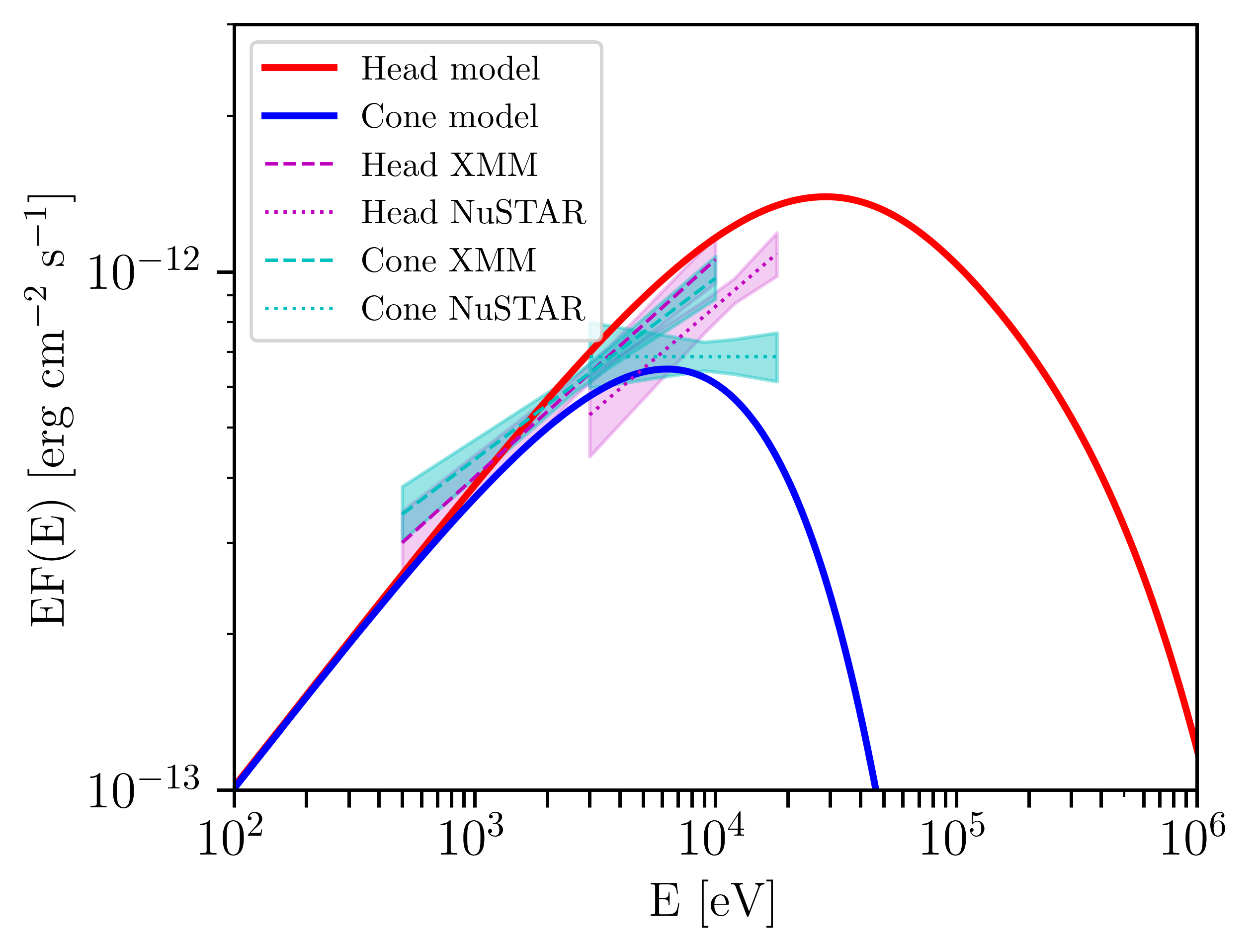}
\caption{Modeled spectrum of X-ray synchrotron radiation in `Head' and `Cone' regions and observational data from \cite{Safi-Harb2022}. The estimated model fluxes of the GHz synchrotron radio emission from the two regions are about a few mJy. } 
\label{synchrotron}
\end{figure}

The exact description of VHE particle transport in the collimated outflow and the surrounding plasma cocoon would require a multiscale modeling of stochastic magnetic fields, which in turn would include different sources of turbulence associated with a complex MHD dynamics with recollimation shocks, shear flows, and cosmic-ray-driven instabilities. Such a model is unfeasible now. Instead, we parameterized the diffusion coefficient of relativistic particles in the system. 

To evaluate the electron distribution function in the downstream flow, we use the advection - diffusion equation, neglecting stochastic acceleration and adiabatic losses:

\begin{eqnarray}
&& \frac{\partial f({\bf r},E)}{\partial t} + {\bf v_{\alpha}(r)}\frac{\partial f({\bf r},E)}{\partial r_\alpha} = \nonumber \\ &=& \frac{\partial}{\partial r_\alpha} D_{\alpha \beta}({\bf r},E)\frac{\partial f({\bf r}, E)}{\partial r_\beta}  - \frac{\partial [a({\bf r},E) f({\bf r},E)]}{\partial E}.
\end{eqnarray}

Here $D_{\alpha \beta}({\bf r},E)$ is the diffusion coefficient,  $a({\bf r},E)=\frac{4\sigma_T}{3m_e^2c^3}\left(B({\bf r})^2/8\pi + U_{ph}\right)E^2$ is the electron synchrotron and Compton loss rates, $\sigma_T$ is the Thompson cross section, $B$ is the magnetic field and $U_{ph}$ is the energy density of the photon field.

\begin{figure}[h!]
\includegraphics[scale=0.4]{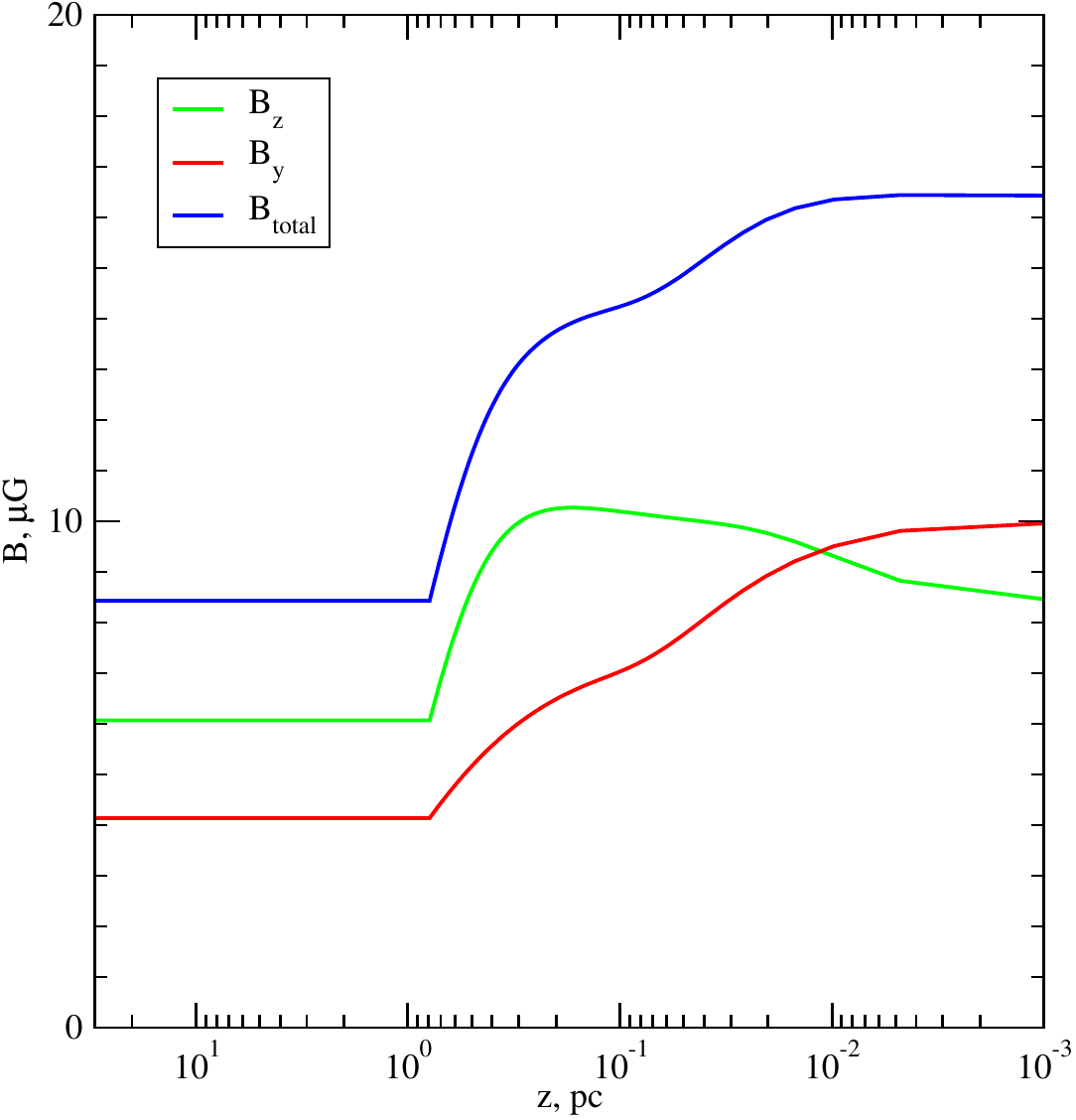}
\includegraphics[scale=0.4]{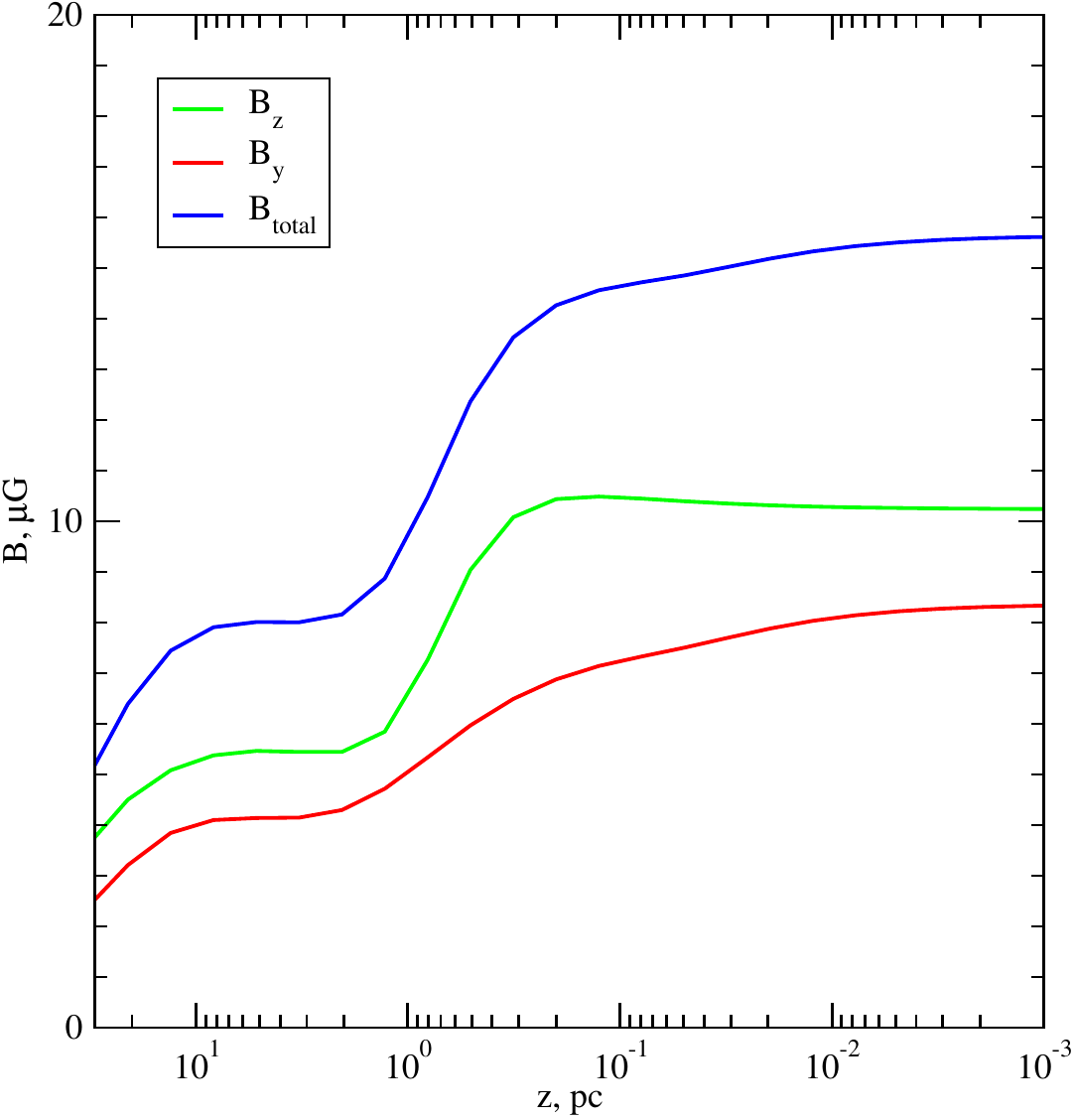}

\caption{\label{fig:Bprof} The upper panel shows the downstream rms magnetic field spatial profile approximations for model 1 described in the text, while
the lower panel shows it for model 2. }

\end{figure}

We simulated a set of models with different assumptions about the diffusion coefficient of VHE particles.   If $D < 3\times 10^{28} \diff$, which is about 100 times larger than the Bohm diffusion coefficient at 100 TeV, the advection dominates the transport along the X-ray jet. Then, in a stationary regime, one can solve this equation by calculating the energy losses of the single electron during its advection into the downstream.

\begin{equation}
    E(z) = \frac{E\left(0\right)}{1+E\left(0\right)k(z)}
\end{equation}

where $k(z) = \int_0^z \frac{4\sigma_T}{3m_e^2c^3}\left(B(z)^2/8\pi + U_{ph}\right) \frac{dx}{v(z)}$. 
The photon field is taken as cosmic microwave background radiation, scattering on all other possible components is suppressed due to the Klein-Nishina regime in the considered energy range (electrons with energies higher than 100 TeV). The turbulent magnetic field was established according to Section \ref{sec:CorSh}. We used two models of the magnetic
field. In the first model, we calculate
only a small-scale turbulent magnetic field. 
In this model, we assume that the field decay stops at some distance
from the shock, so the magnetic field remains constant onward. 
In our calculation, we assume that the critical value at which the field decay stops is $8$ $\mu G$. In Model 2, we assume a superposition of the simulated short- and long-scale magnetic fields calculated
for $l_{*}=3\cdot10^{17}$~cm and $l_{*}=3\cdot10^{19}$~cm. 
Using the results of MHD magnetic field simulations discussed in Section \ref{sec:CorSh}, we constructed a smooth analytical approximation for square-averaged magnetic fields, neglecting small-scale fluctuations, as was done in \cite{2024PhRvD.110b3041B} for Tycho SNR. The averaging
is performed on slices of the simulated MHD datasets at fixed coordinates $z$. 
The square average magnetic field profiles for models 1 and 2
are shown in Fig.~\ref{fig:Bprof}. 
The approximation has a different
functional dependence on the distance from the shock for normal ($B_{\parallel}$) and transverse (${B_{\perp}}$) field components.

Thus, in the advection model, the electron distribution at a distance $z$ from the shock in the downstream can be evaluated as

\begin{equation}
    f(z,E)=f\left(0,\frac{E}{1-k(z)E}\right)\frac{1}{\left(1-k(z)E\right)^2}
\end{equation}

where $f(0,E)$ is the distribution function of electrons at the shock front, shown in Figure~\ref{pdf_shock_Q_esc}. 
The modeled level of synchrotron radiation derived with the electron-to-proton ratio fixed to be 3.3$ \times 10^{-3}$ at 1 GeV described above fits well with the observed data.   The energy spectrum of the `Head' and `Cone' regions in the short-scale Bell's turbulent field model with characteristic length $l_\ast = 3\times10^{17}~\rm{cm}$, maximum field on the front $15~\rm{\mu G}$ and constant level $8~\rm{\mu G}$, constant velocity $v = 10.3\times10^8~\rm{cm/s}$ (consistent with Monte Carlo simulation), and observational data \cite{Safi-Harb2022} are shown in Fig.~\ref{synchrotron}. The modeled spectra are consistent with the power-law functions obtained in the observations.

\begin{figure}[h!]
\includegraphics[scale=0.65]{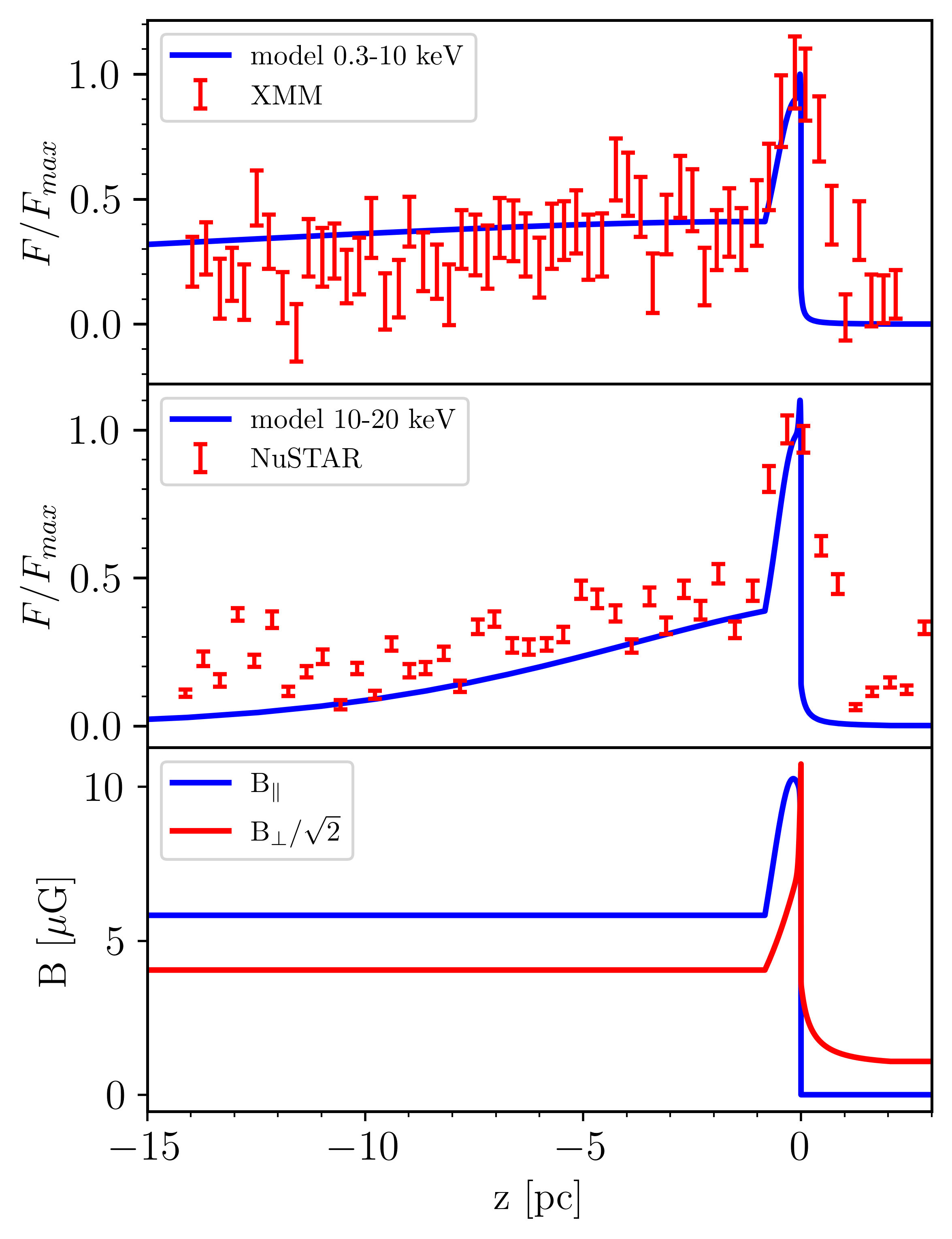}
\caption{Spatial profiles of synchrotron emission in e1 region in the energy range 0.3-10 keV predicted by the model in comparison with the XMM data \cite{Safi-Harb2022} (top panel), in the energy range 3-30 keV and the NuSTAR data \cite{Safi-Harb2022} (middle panel), and the profile of magnetic field used in the model (bottom panel).} 
\label{profile_e1}
\end{figure}

\begin{figure}[h!]
\includegraphics[scale=0.65]{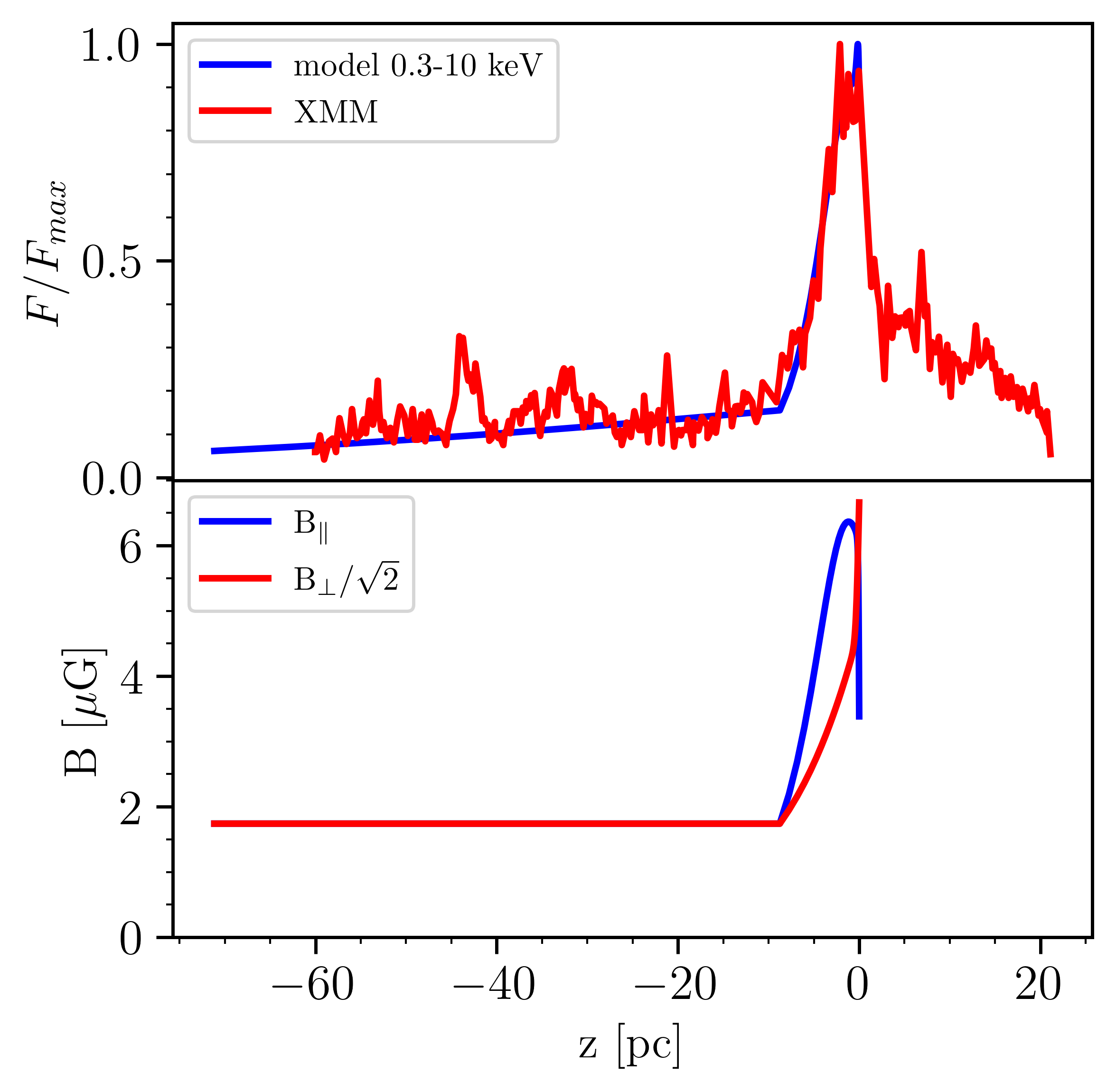}
\caption{Modeled spatial profile of synchrotron radiation in e2 region in energy range 0.3-10 keV and XMM data \cite{Brinkmann2007} (top panel), and profile of magnetic field used for modeling radiation (bottom panel). } 
\label{profile_Brinkmann}
\end{figure}

Spatial profile of the X-ray luminosity for the same model is shown in Figure~\ref{profile_e1}. It is also consistent with observations. The profile for the `e2' region is described in \cite{Brinkmann2007}, and was studied only by XMM. To explain this case, we use a turbulent field with the maximum value of $9~\rm{\mu G}$, constant level of $3~\rm{\mu G}$, characteristic length $2\times10^{18}~\rm{cm}$ and velocity $9.3\times10^8~\rm{cm/s}$, results of simulation are shown in Fig.~\ref{profile_Brinkmann}.

\begin{figure}[h!]
\includegraphics[scale=0.45]{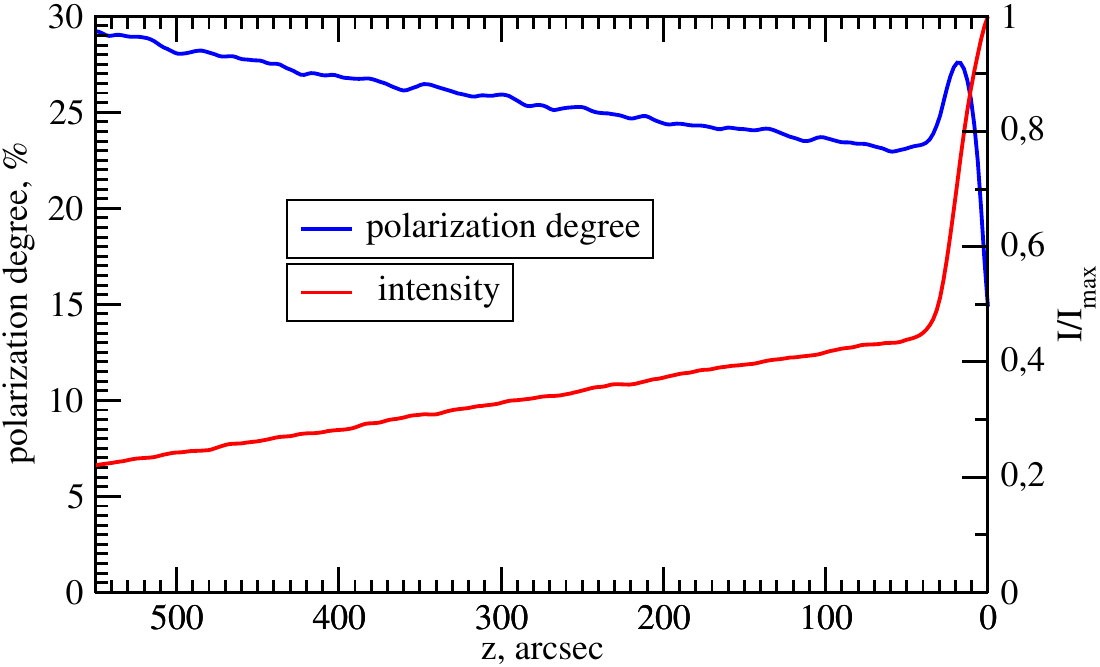}
\includegraphics[scale=0.45]{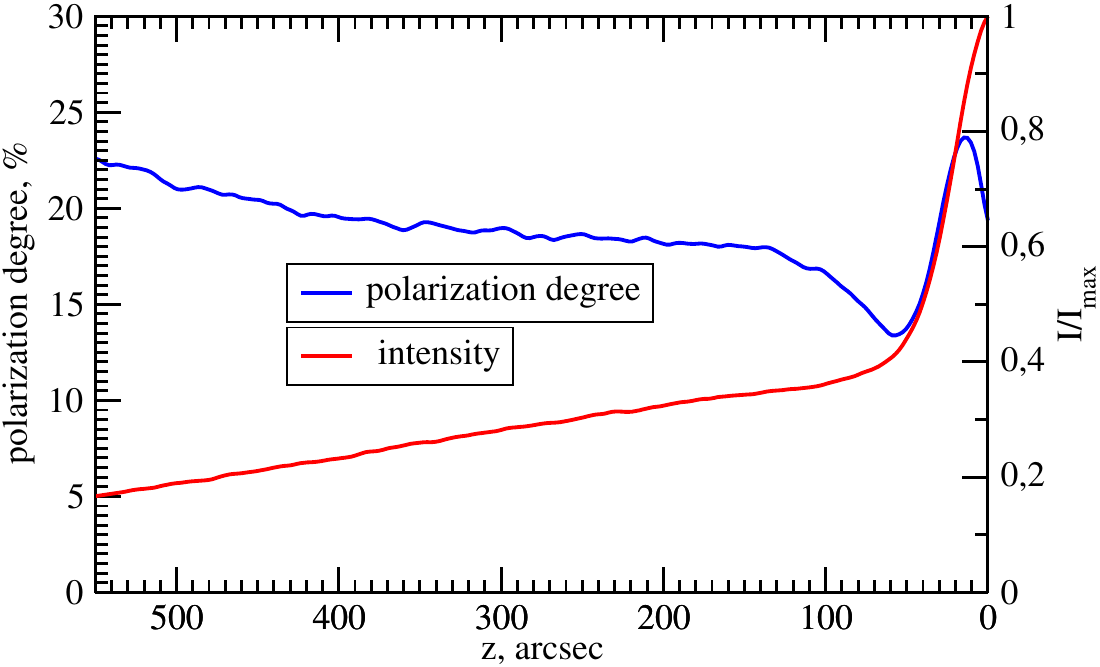}

\caption{\label{fig:polar} The upper panel shows polarization degree and emission simulated for
model 1 and the lower panel for model 2. IXPE PSF is taken into account.}
\end{figure}

The models of the magnetic field described above were used to predict
Stokes parameters I, U, Q, and polarization degree profiles of the jet emission in
the downstream region after the shock, using expressions for synchrotron emission from \citep{Ginzburg1965}. 
To model integration over the line
of sight (Ox axis), the averaging over the Gaussian probability distribution function
of the $B_{y}$ and $B_{z}$ projections was done. We take into
account the PSF of IXPE (\citep{Fabiani_2014ApJS..212...25F} ) and assume a photon energy of 3~keV and no unpolarized thermal emission.  We also assume that the magnetic field realization
over the line of sight forms a complete statistical ensemble. This
is not absolutely true, so the real values can fluctuate around simulated
ones. However, taking into account IXPE PSF, the amplitude of such fluctuations is expected to be lower 
due to additional averaging. This method was used in the work \cite{2024PhRvD.110b3041B} for Tycho SNR.
The model results are shown
in Fig.~\ref{fig:polar}.

\begin{figure}[t!]
\includegraphics[scale=0.65]{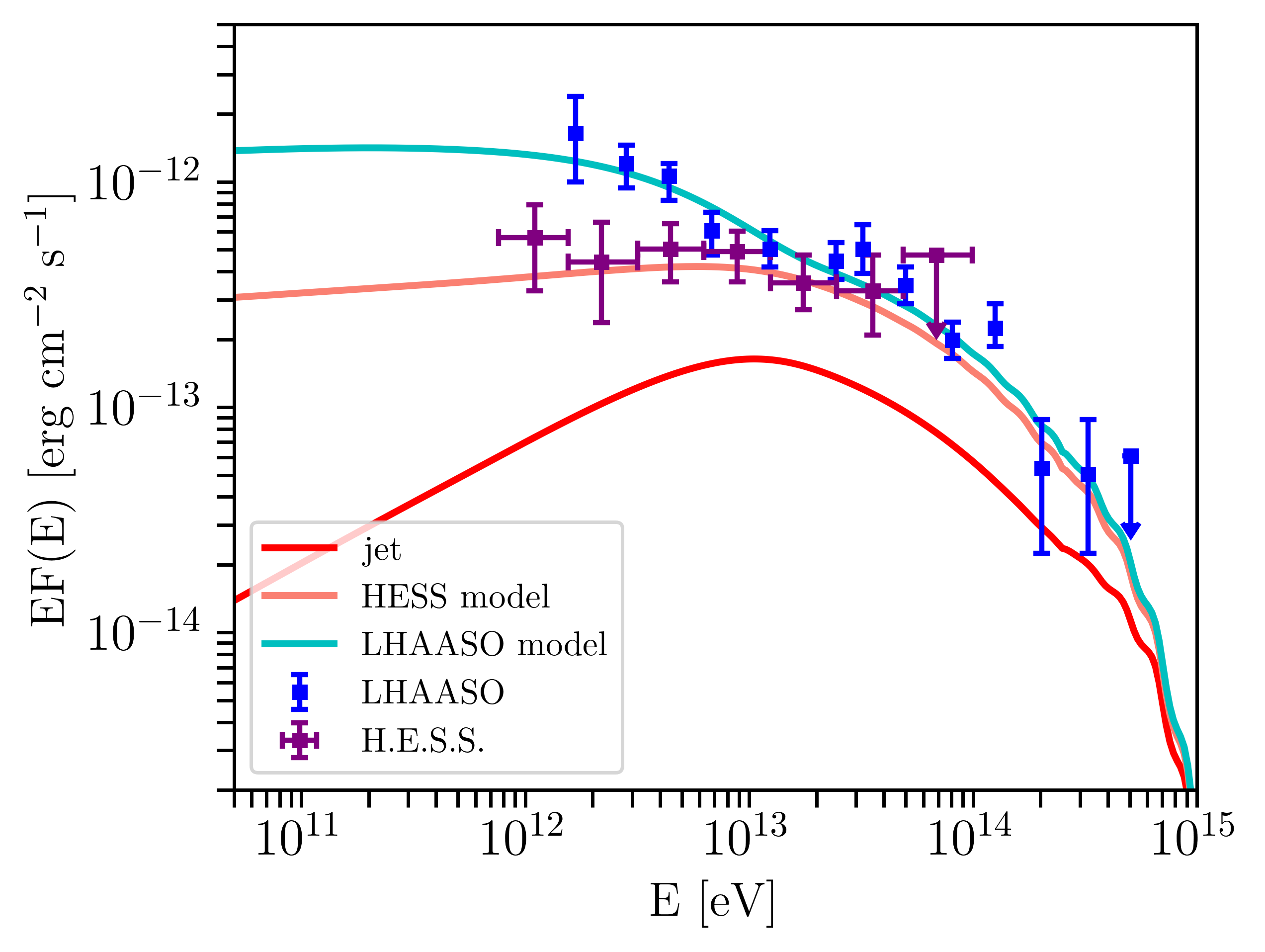}
\caption{Modeled spectrum of the inverse Compton radiation of W50 nebula and observational data from \cite{HESS2024SS433, LHAASO2024SS433}. }
\label{compton}
\end{figure}

\section{Gamma-Ray emission model}
\label{sec:Gamma}

In this Section, we model the gamma radiation of W50, which is observed by H.E.S.S. \cite{HESS2024SS433} and LHAASO \cite{LHAASO2024SS433}. The leptonic model assumes that it is produced via inverse Compton scattering of electrons accelerated by recollimation shocks in the collimated outflow.  

In Fig. \ref{compton}, the red curve shows the gamma radiation produced within the outflow by accelerated VHE electrons as they are advected through the downstream of the first shock to the second one. This corresponds to pure advection of VHE electrons along the X-ray outflow with the turbulent magnetic field and the distribution function of electrons, which were used to simulate the synchrotron X-ray radiation discussed above. The red curve is significantly below the observational data, and it has a different spectral shape at low energies. This is because the radiating electrons in the X-ray jet are in the thin target regime.
On the other hand, the electrons accelerated at the first shock diffuse in the direction transverse to the X-ray outflow. These electrons left the axial X-ray outflow and diffused through the surrounding cocoon, comprising a spatially broad component. If their diffusion coefficient in the cocoon is $\sim 10^{28} \diff$ at 100 TeV, they would radiate gamma-rays in the thick target regime in a wide cocoon of a size $\sim$ 30 pc around the axial jet. Electrons that have flown through the X-ray axial jet with a magnetic field amplified by cosmic ray-driven instabilities at the shock and escape into the cocoon downstream would also radiate in the thick regime. In Fig. \ref{compton}, the gamma-ray spectrum integrated over the wide cocoon is shown by the blue curve, which is in reasonably good agreement with the LHAASO data points taken from \cite{LHAASO2024SS433}. The gamma-ray spectrum integrated over a somewhat narrower region of the cocoon corresponds to the intermediate regime between the thin and the thick targets. It is shown by a brown curve in Fig.~\ref{compton} and is compared with the data reported by H.E.S.S. \cite{HESS2024SS433}.

The LHAASO images above 100 TeV \cite{LHAASO2024SS433} show a more extended emission region compared to the lower energy bands. This morphology can be understood within our diffusive shock acceleration model, where the highest energy particles escape into the shock upstream, where the low magnetic field allows the escaped VHE leptons to form an extended gamma-ray emission region by the inverse Compton process.

In our model, $\sim$ 30 $\%$ of the electrons accelerated at the shock front had escaped from the jet and diffused in the cocoon. This allows us to explain the observational data from LHAASO as shown in Figure \ref{compton}. H.E.S.S. data fall between the model for LHAASO and radiation from the jet itself, which can be explained by a smaller region size, observed by H.E.S.S.

There are recent studies of the acceleration of PeV protons in microquasars \citep{2024A&A...691A..93A,2024arXiv241108762P,2025arXiv250510620K}.  A spherically symmetric semi-analytical model of particle acceleration to energies of several PeV by the termination wind shock was discussed in \cite{2024arXiv241108762P}. The acceleration of particles in the shear flows of jet-cocoon-structured microquasars is suggested in \citep{2025arXiv250620193Z}. Shear acceleration in the model requires the injection of particles pre-accelerated up to TeV regime energies. Hadronic interactions of accelerated protons are producing a substantial fraction of the observed VHE gamma-ray emission in the models \citep{2024arXiv241108762P,2025arXiv250510620K}. 

The recollimation shocks in the diffusive shock acceleration model considered here efficiently accelerate protons up to PeV energies, as is seen in Fig. \ref{pdf_shock_Q_esc}.
In our model, $\gsim$ 10\% of the shocked jet power is transferred to accelerated protons while the VHE emission comes from the inverse Compton radiation of VHE electrons accelerated at shocks, producing polarized synchrotron X-ray emission of the extended X-ray jet. The leptonic gamma-ray emission is produced in the thick target regime within the cocoon surrounding the extended X-ray jet. 
   The VHE gamma rays can be produced by hadronic interactions of the protons with ambient matter. The very low number density of the plasma in the extended jet and in the surrounding cocoon would require a very small diffusion coefficient for the accelerated VHE protons to contribute to the detected gamma-ray emission from the cocoon, given by the gamma-ray significance maps of H.E.S.S. and LHAASO. Therefore, most of the accelerated protons would escape the accelerator in the SS433/W50 extended X-ray jet and then contribute to the gamma-ray emission from the denser gas in the vicinity of W50 and supply the galactic cosmic-ray population.      

\section{Summary}

We present the results of detailed modeling of nonthermal processes in the extended outflow produced due to supercritical accretion onto a black hole in SS433/W50. The model relies on the minimalist scenario of the outflow origin, which supposes an interaction of two outflows from the complex source - a nearly isotropic wind and a more collimated outflow \citep{Churazov}. Strong shock waves are present in the region where the collimated outflow is passing through the termination surface of the isotropic wind. MHD modeling of interacting outflows with the PLUTO code for the wind and outflow powers $\gsim 10^{39} \ergs$ shows a strong recollimation shock at a location which is compatible with the beginning of the observed extended non-thermal X-ray and VHE gamma-ray features. To simulate the spectra of protons and electrons accelerated at the strong shock by the diffusive shock acceleration mechanism, we used the nonlinear Monte Carlo model, which accounts for turbulent magnetic field amplification at the shock due to cosmic-ray-driven instabilities. The electric current of the accelerated VHE protons escaping the accelerator is producing strong magnetic fluctuations by Bell's instability in the upstream of the strong shock. The transport of turbulence through the shock upstream, its transformation during the passage through the shock front, and its evolution in the downstream were simulated with the PLUTO MHD module. Then, the spectra and spatial distribution of the nonthermal emission in the shock vicinity that is dominated by the synchrotron and inverse Compton emission of the accelerated electrons are derived. 

In the model, relativistic protons and electrons are accelerated up to the PeV energy regime, and a sizable fraction of the shock ram pressure (i.e., the outflow power) can be transferred into VHE protons. The magnetic field amplification at the shock is mostly due to cosmic ray-driven instabilities, and the magnetic field is highly fluctuating, as is the case for the efficient diffusive shock acceleration mechanism. The magnetic turbulence produced at the shock is decaying farther downstream. The synchrotron radiation of the accelerated VHE electrons in the magnetic field downstream from the shock is in the X-ray band. The model allows one to explain both the spatial profiles and the spectra of the nonthermal X-ray emission detected by XMM-Newton and NuSTAR \citep{Brinkmann2007,Safi-Harb2022}. Moreover, an important prediction of the models is the polarized X-ray emission with the predominant direction of the magnetic field along the extended eastern jet, such as detected recently by the IXPE observatory  \citep{IXPE2024SS433}. This result is consistent with our model, which relies on the diffusive shock acceleration mechanism described above. The magnetic field is turbulent in the synchrotron emitting region in the shock downstream. However, in sufficiently strong shocks, the turbulence is anisotropic with the dominant direction of the fluctuating magnetic field along the post-shock flow direction, which produces the polarized X-rays \citep{2024PhRvD.110b3041B}. 

VHE gamma-ray emission along the extended X-ray jet starting from the `e2' region was reported by the H.E.S.S. collaboration \citep{HESS2024SS433}, and photons above 100 TeV from a wide vicinity of SS433 were detected by the LHAASO observatory \citep{LHAASO2024SS433}. We have shown that the detected VHE emission can be attributed to the inverse Compton radiation of electrons accelerated at the strong shocks in the extended jet. Less than a percent of the jet power (imparted at the shock to the accelerated electrons) is sufficient to account for the observed VHE radiation. Much larger power is going into the accelerated PeV regime nuclei. These protons are escaping the vicinity of the outflow without any significant energy losses because of the very low plasma density here. The escaped nuclei may interact with the HI gas, which may have a mass up to 10$^5 \Msun$ in the expanding shell of about 70 pc radius \citep{Lockman_W50_HI_07}, providing an extended VHE emission which depends on the diffusion coefficient of PeV regime particles in the region which we discuss below.

Strong shocks in the vicinity of the `e2' region and the second consequent recollimation shock accelerate protons to $\sim$ PeV maximal energies with the efficiency of $\gsim$ 10\% of the outflow power. That is, from the total power $\sim 10^{39} \ergs$ of the extended east outflow modeled, about $\sim 1.7 \times 10^{38} \ergs$ is the kinetic power passing through the shock. The power transferred to the protons accelerated above 50 TeV at the shock is $\sim 3 \times  10^{37} \ergs$, while $\gsim  10^{35} \ergs$ was transferred to the electrons. A similar power partition happens at the second shock. The eastern extended collimated outflow may provide about $5 \times  10^{37} \ergs$ to very high-energy cosmic rays. Thus, both the Eastern and Western outflows together may contribute a power $\sim  10^{38} \ergs$ to CR protons above 50 TeV. 
As concerns the maximal energy of the accelerated protons, satisfactory fits to the X-ray and VHE gamma-ray emission of the extended jet can also be obtained with
slightly different parameters of the first shock. For example, if the free escape boundary is at $L_{FEB}=6$ pc and  $B_{turb0}=3.0$~$\mu$G then the protons can be accelerated to several PeV, and the electrons to energies of about 1 PeV. In this case, the synchrotron radiation from the `head' region extends to MeV photon energies with the flux $\sim 10^{-12} \enf$ at 1 MeV, which can be detected with future MeV missions, thus testing the maximal energies of accelerated protons in our model.  

The model of extended X-ray jets of SS 433 discussed above allowed us to explain the very high energy gamma-ray data obtained by the H.E.S.S. and LHAASO observatories with just leptonic emission. The protons accelerated to PeV regime energies and carrying up to 10\% of the jet power may contribute to the gamma-ray emission through proton-proton interactions. The cooling time of a very high energy proton is $\gsim 10^{15}/n$ s, where the matter number density is measured in $\cmc$.  Given the low number density of the matter both in the jet and the cocoon, much less than 0.1\% of the proton luminosity is radiated in hadronic interaction gamma-rays during the system age of $\sim 30,000$ yrs, which is less than the inverse Compton emission of electrons. If PeV protons escaped from the accelerator would propagate with the diffusion coefficient $> 3\times 10^{29} \diff$ extrapolated from the GeV-TeV range (see e.g. \citep{2007ARNPS..57..285S}) then they can interact with the dense molecular clouds in the vicinity of $\gsim$ 300 pc from the source. On the other hand, the plasma instabilities accompanying the CRs escaping a powerful source can greatly reduce the diffusion coefficient (by a few orders of magnitude compared to the global value used above) on scales of tens of parsec around it. 

The exact model of the CR escape from a powerful CR source is a kinetic multiscale non-linear problem (see e.g. \cite{2008AdSpR..42..486P,bbmo13} and the recent review \citep{2025FrASS..1111076M}) and thus it is not yet available. Instead, a very simplified back-of-the-envelope  estimations of the CR proton propagation scale can be made with a toy model where the quasi-linear growth rate of the resonant instability \citep{1982A&A...116..191M} is balanced by the magnetic turbulence cascading at any given point. The growth of magnetic fields occurs on the scale of the gyroradius of a CR particle $r_{g}$. 
Outside the accelerator, the CR distribution function obeys the equation: 
\begin{equation}\label{diff_cr}
\frac{1}{r^{2}}\frac{d}{dr}\left(r^{2}D\frac{df_{cr}}{dr}\right)=0,
\end{equation}
where $D(r,p)$ is the diffusion coefficient to be evaluated, $f_{cr}(r,p)$ is the particle distribution function. The momentum distribution of the CRs escaping the shock as can be seen in Fig. \ref{pdf_shock_Q_esc} is rather narrow. This simplifies further analysis. In this case, the CR-generated turbulence can be characterized by a single scale  $L=r_{g}$ and the magnetic field amplitude $\delta B$.

The CR proton gradient $\frac{dP_{cr}}{dr}$ which determines the quasi-linear growth rate of the resonant instability can be expressed from the solution of Eq.(\ref{diff_cr}): 
\begin{equation}\label{gradPcr2}
\frac{dP_{cr}}{dr}=-u_{2}\frac{4\pi}{3}\int_{p_{min}}^{\infty}vpf_{d}(p)p^{2}dp\frac{S}{4\pi r^{2}D(r,p)},
\end{equation}
where $S$ is the area of the shock from which CR escaped, $f_{d}(p)$ is the particle distribution function in the shock downstream, $u_{2}$ is the plasma speed just in the downstream, $v\approx c$ is the particle velocity. In the source vicinity, where the turbulent magnetic field produced  by the escaping CRs exceeds the ambient field in the cocoon, the diffusion coefficient is quasi-isotropic and can be approximated by the Bohm diffusion coefficient $D_B$:

\begin{equation}\label{Diff_coeff}
D(r,p) = D_B=\frac{r_{g}v}{3}.
\end{equation}
where $r_{g} = cp/e\delta B(r)$.

Then, from the balance equation between the turbulence growth and decay rates    
\begin{equation}\label{gradPcr}
\frac{dP_{cr}}{dr}=-\frac{\left(\delta B\right)^{2}}{4\pi L}.
\end{equation}
one can estimate the amplitude of the amplified turbulent magnetic field as  
\begin{equation}\label{B_value}
\delta B(r) =\left(\frac{u_{2}}{c}4\pi\int_{p_{min}}^{\infty}vpf_{d}(p)p^{2}dp\frac{S}{r^{2}}\right)^{\frac{1}{2}}.
\end{equation}

For the parameters of the first shock, 
\begin{equation}\label{B_value_fs}
\delta B\approx  \left(\frac{10 pc}{r}\right) \mu G,
\end{equation}
 The diffusion approximation is valid at distances R which are well above the particle mean free path. 
 From Eq.(\ref{B_value_fs}) it follows that this condition is fulfilled for PeV protons.
 Thus, one can estimate that the accelerated protons can reach a distance of about 200 pc in 30,000 yrs, while particles of energy below 100 TeV will be confined within the nebula.

Since a substantial fraction of the supercritical accretion disk power can be carried by the collimated extended jet, this source may contribute substantially to the PeV regime cosmic ray sea during its supercritical accretion cycle. While the statistics of the ultraluminous accreting sources and their duty cycles are not certain yet, some very rough estimates can support this point. 

Indeed, the total luminosity of the comic ray sources of energies above PeV can be estimated as $\sim 2 \times 10^{38} \ergs$ from the extrapolation of the scaling relation presented in \citep{2019PhRvD..99f3012M} for the GeV-TeV range. The scaling in \citep{2019PhRvD..99f3012M} relies on the measured ratio of secondary and primary CR fluxes and the GeV-TeV CR source distribution, which are likely core-collapse supernovae. Certainly, the assumption that the PeV regime CRs would follow the same propagation model as constructed for the GeV-TeV range is not justified if the PeV CR sources are microquasars or other rare types of sources, and the total luminosity given above can be regarded as an order-of-magnitude estimate at best.    

Assuming that the number of microquasars in the Galaxy is about 10, the LHAASO team \citep{LHAASO2024SS433} estimated the expected fluxes of PeV protons at Earth within the leaky box model, which were consistent with the observed values. The spatial distribution of PeV regime CRs in the Galaxy is under study with the current sensitive Cherenkov observatories. However, the low observed dipole anisotropy of the galactic cosmic ray PeV protons of $\sim 10^{-3}$ \citep{2025ApJ...981..182A} places serious constraints on the CR propagation model with a small number of sources.

\begin{acknowledgments}
The authors thank two reviewers for carefully reading this paper and for reports
that helped us clarify our results.
X-ray polarization simulations of extended jets by Yu.~U. were supported by the baseline project FFUG-2024-0002 at the Ioffe Institute. RMHD modeling of the structure of W50/SS433 by V.R. was supported by RSF grant No. 25-72-20007 (Multiscale nonlinear models of astrophysical sources of high energy radiation). Monte Carlo modeling by S.O. was performed at the Supercomputing Center of the Peter the Great Saint-Petersburg Polytechnic University and the Joint Supercomputer Center.
IK acknowledges the support by the COMPLEX project from the European Research Council (ERC) under the European Union’s Horizon 2020 research and innovation program grant agreement ERC-2019-AdG 882679.
\end{acknowledgments}

\section*{References}

\bibliographystyle{apsrev}
\bibliography{bibliogr}
\end{document}